\documentclass[twocolumn,english,aps,prl,floatfix,showkeys]{revtex4-1}
\usepackage[utf8]{inputenc}
\usepackage{color}
\usepackage{amsmath}
\usepackage{amssymb}
\usepackage{graphicx}
\usepackage{esint}
\usepackage[english]{babel}
\usepackage[sort&compress]{natbib}
\usepackage[unicode=true, pdfusetitle, bookmarks=true, bookmarksnumbered=false, bookmarksopen=false, breaklinks=false, pdfborder={0 0 0}, backref=false, colorlinks=true, citecolor=blue, filecolor=black, linkcolor=black, urlcolor=blue] {hyperref}
\usepackage{textcomp}
\usepackage{multirow}
\usepackage{mciteplus}
\usepackage{babel}
\begin{document}

\title{Giant-spin nonlinear response theory of magnetic nanoparticle hyperthermia: a field dependence study}

\date{\today}

\author{M. S. Carrião}

\email{Corresponding author: mscarriao@ufg.br}

\author{V. R. R. Aquino}

\affiliation{Instituto de F\'{i}sica, Universidade Federal de Goiás, 74690-900,
Goiânia-GO, Brazil}

\author{G. T. Landi}

\affiliation{Departamento de Ciências Naturais e Humanas, Universidade Federal
do ABC, 09210-580, Santo André-SP, Brazil}

\author{E. L. Verde}

\affiliation{Instituto de Ciências Exatas e da Terra,
Universidade Federal de Mato Grosso, 3500, Pontal do Araguaia-MT, Brazil}

\author{M. H. Sousa}

\affiliation{Faculdade de Ceilândia, Universidade de Bras\'{i}lia, 72220-140,
Bras\'{i}lia-DF, Brazil}

\author{A. F. Bakuzis}

\affiliation{Instituto de F\'{i}sica, Universidade Federal de Goiás, 74690-900,
Goiânia-GO, Brazil}

\begin{abstract}
Understanding high-field amplitude electromagnetic heat loss phenomena
is of great importance, in particular in the biomedical field, since
the heat-delivery treatment plans might rely on analytical models
that are only valid at low field amplitudes. Here, we develop
a nonlinear response model valid for single-domain
nanoparticles of larger particle sizes
and higher field amplitudes in comparison to linear
response theory. A nonlinear magnetization expression and a generalized
heat loss power equation are obtained and compared with the exact
solution of the stochastic Landau-Lifshitz-Gilbert equation assuming
the giant-spin hypothesis. The model is valid within the hyperthermia
therapeutic window and predicts a shift of optimum particle size and
distinct heat loss field amplitude exponents. Experimental hyperthermia
data with distinct ferrite-based nanoparticles,
as well as third harmonic magnetization data supports the nonlinear
model, which also has implications for magnetic particle imaging and
magnetic thermometry.
\end{abstract}


\maketitle

\section{Introduction}

The response of nanomaterials to alternating electromagnetic fields
is of great importance nowadays in the biomedical field, where new
approaches to treat diseases are under development. One of the most
innovative and important applications is related to heat delivery
through the interaction of nanomaterials with electromagnetic fields.
This heat delivery method can be used to release drugs\citep{HoareNL09},
activate biological processes\citep{HuangNatNano10,TorayaBrownNNBM14,KobayashiNanomed14}
and even treat tumors\citep{RodriguesIJH13,DenisIJH13,HilgerIEEEProcNano05,MaierHauffJNeuro11,GilchristAnnSur57}.
Indeed, using Maxwell's equations and the first law of thermodynamics
one finds that the heat loss per unit volume per cycle is given by
\begin{equation}
\dfrac{1}{V}\oint\limits _{\mathrm{cycle}}\delta Q=\int\vec{E}\cdot\vec{J}\mathrm{d}t-\oint\limits _{\mathrm{cycle}}\vec{P}\cdot\mathrm{d}\vec{E}-\oint\limits _{\mathrm{cycle}}\mu_{0}\vec{M}\cdot\mathrm{d}\vec{H},
\label{GenHeatLoss}
\end{equation}
where $V$ is the nanomaterial volume, $Q$ is the heat loss, $\vec{E}$
the electric field, $\vec{J}$ the free volumetric density current,
$\vec{P}$ the electric polarization, $\mu_{0}$ the vacuum magnetic
permeability, $\vec{M}$ the magnetization and $\vec{H}$ the magnetic
field. The first term in equation \eqref{GenHeatLoss} correspond
to the ``free-current'' loss, whereas the last two describe dielectric
and magnetic losses.

\begin{figure}[htb]
\centering \includegraphics[width=1\columnwidth]{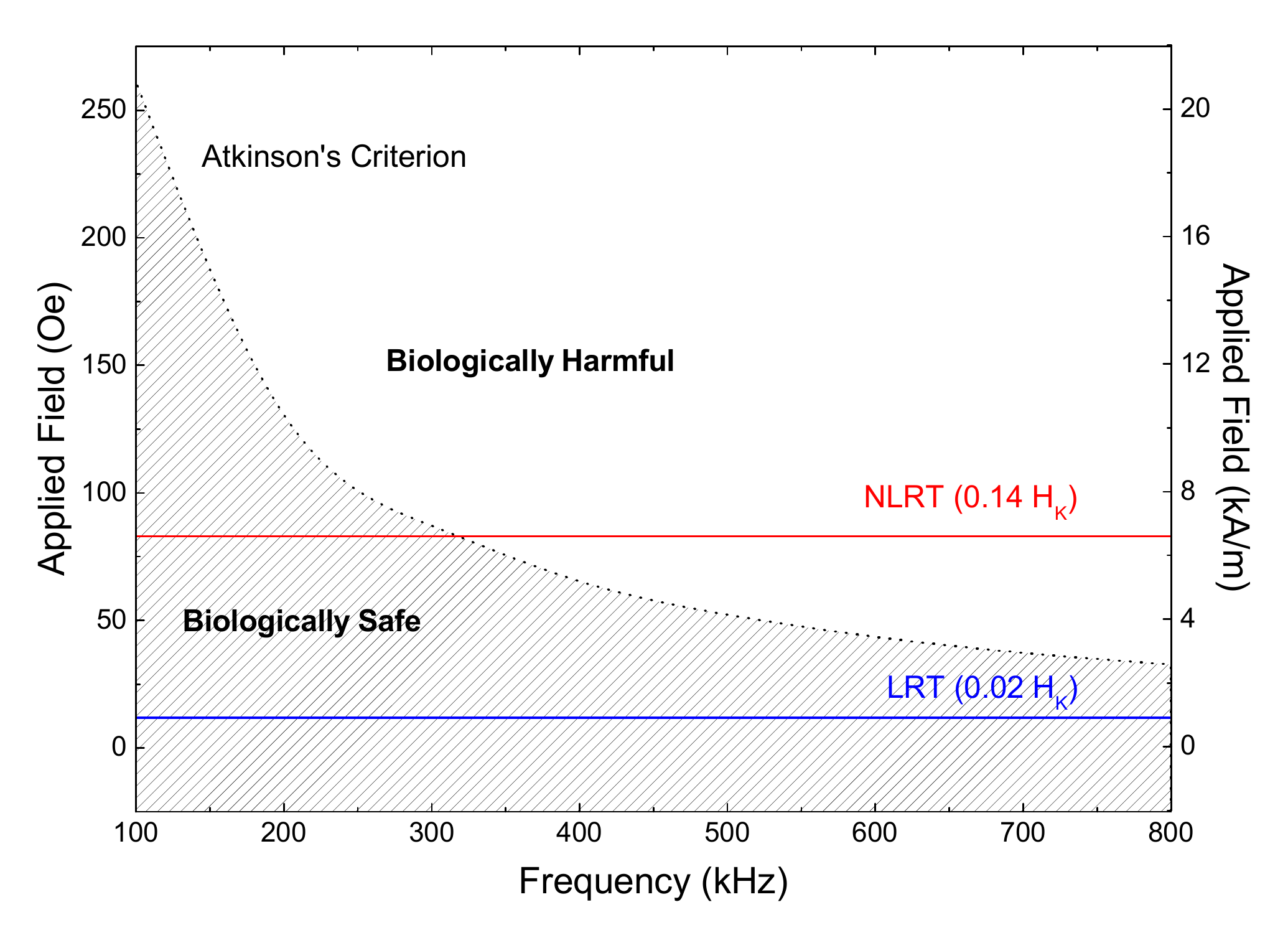} \caption{(Color online) The figure shows the calculated biological critical
field according to Atkinson's criteria scaled for a coil radius of
$0.035~\mathrm{m}$ (expected for small tumors in
breast \citep{EtheridgeABE13}), which results in a $Hf<20.78\times10^{5}~\mathrm{kA/(m}\times\mathrm{s)}$
. Human treatment can only occur below this field. The LRT limit is
calculated assuming $H<0.02H_{K}$ for a particle diameter of $15~\mathrm{nm}$,
$M_{S}=270~\mathrm{emu/cm^{3}}$, $K_{ef}=8\times10^{4}~\mathrm{erg/cm^{3}}$,
$T=300~\mathrm{K}$, $\alpha=0.05$ and $\rho=5~\mathrm{g/cm^{3}}$.
The nonlinear response critical field for our model (NLRT) corresponds
to the solid line.}
\label{figure1} 
\end{figure}

Moreover the ``free-current'' loss term has an important impact
on the biomedical application since it is related to a biological
constraint. This term states that the frequency ($f$) and magnitude of the
alternating magnetic fields need to be lower than a critical value
in order to inhibit possibly harmful ionic currents in the patient's
body\citep{HilgerIEEEProcNano05}. For instance, for a frequency of
100 kHz the maximum field amplitude is in the order of $20.8~\mathrm{kA/m}$
($261~\mathrm{Oe}$) for a single air-core coil radius
of $0.035~\mathrm{m}$ (expected dimension for breast cancer application\citep{EtheridgeABE13}).
Note that this value is higher than the one usually
reported (order of $4.9~\mathrm{kA/m}$) only because
the estimation of Atkinson used a coil radius of $0.150~\mathrm{m}$. Since the
free current loss is proportional to the square of the distance from
the coil axis, an estimation of the critical field for a given coil
radius ($r$) might be obtained from $Hf<(0.150/r)*4.85\times10^{5}~\mathrm{kA/(m}\times\mathrm{s)}$. Figure \ref{figure1} shows the biological critical field as a function
of field frequency in the usual therapeutic hyperthermia range using
Atkinson's criteria\citep{HilgerIEEEProcNano05,AtkinsonIEEEBME84,EtheridgeABE13},
which indicates that the higher the frequency the lower is this field
(the parameters used to generate the curve are presented in the figure
captions).

On the other hand, the last terms of Eq. \eqref{GenHeatLoss}, which
represent hysteretic losses, has been the subject of analytical models
within the Linear Response Theory (LRT) and was used to estimate optimal
particle size, understand particle-particle interaction effects and
maximum heat generation for hyperthermia\citep{DebyePolMolec29,HergtIEEETM98,RosensweigJMMM02,BranquinhoSR13}.
Curiously, most LRT studies from the literature do not discuss a fundamental
limitation of the model, namely, that it is only valid at the low
particle size range and low field amplitudes. 

In Fig.\ref{figure1} we show the range of validity of the LRT, which,
as can be seen, is far below the typical fields used for hyperthermia
studies. There are several suggestions for identifying this limit.
For example, Carrey et al.\citep{CarreyJAP11} found that the hysteresis
area for a longitudinal case (field applied parallel to the easy axis
- see Fig. 5(g) of Ref. \citep{CarreyJAP11}) deviates from the LRT
for $\xi\leq0.2$ ($\xi=\mu_{0}M_{S}VH/k_{B}T$ - where $M_{S}$ is
the saturation magnetization, $k_{B}$ is the Boltzmann constant and
$T$ is the temperature), which suggests that this model can only
be applied for particles below a critical size. Alternatively,
Verde et al. suggested that deviations occurs for fields $H<0.02H_{K}$($H_{K}$
is the anisotropy field, that for uniaxial case is
$H_{K}=2K/\mu_{0}M_{S}$ with $K$ the anisotropy
constant)\citep{VerdeAIPAdv12,VerdeJAP12}. It is important to emphasize
that experimentally, it is possible to determine
if one is still in the linear regime or not, by verifying if the the
heating efficiency (also known as specific loss power - SLP) scales
quadratically with the field. Throughout this manuscript,
when discussing the theoretical models, low magnetic fields mean values
within the LRT range. In addition, in Fig. \ref{figure1} we also
include an estimation of the range of validity of the nonlinear response
model (NLRT) developed in this work, which will be
shown later in the manuscript to be $H<0.14H_{K}$. This corresponds
to a 7-fold increase in the range of field validity in comparison
to the LRT definition used above. The result suggests that the model
may be useful for biomedical applications, in particular for magnetic
hyperthermia.

In the subject of heat loss, the term ``nonlinear'' has been used
in a variety of ways. For instance, nonlinear dielectric effects have
been related to the correlation of distinct relaxation times\citep{RichertPRL06}.
In this case, a superposition of Debye processes is used, which predict
heat loss scaling with the square of field amplitude. Conversely,
for relaxor ferroelectric materials a nonlinear polarization term
is included in the dynamic response equation\citep{GlazounovPRL00}.
Such approach allowed the authors to investigate the third harmonics
of the relaxor. On the other hand, in magnetic materials, nonlinear
response is investigated using the stochastic Landau-Lifshitz-Gilbert
(SLLG) equation\citep{VerdeAIPAdv12,VerdeJAP12,GarciaPalaciosPRL00,LandiJAP12c}.
In this case, thermal fluctuations are addressed using the Brown's
approach\citep{BrownPR63}, where the giant spin hypothesis allow
one to use the Fokker-Planck equation to the magnetic moment orientational
distribution function. One can then show that this leads to an infinite
hierarchy of equations, which can be solved numerically to find the
magnetic moment response of the nanoparticle \citep{DejardinJAP09,LandiJAP12b,PoperechnyPRB10,LandiPRB14}.
The method is valid for any field amplitude, but due to its
mathematical complexity, it does not yield simple
analytical expressions that could be useful in the applied field.

Indeed, the field and frequency-dependence of heat loss
in magnetic materials have been attracting the attention
for a long time due to technological applications 
\citep{BertottiIEEETM88}. In general, the loss in magnetic materials can
show several contributions, spanning from eddy currents
(that scales with $f^2H^2$), anomalous eddy current con-
tributions (due to complex domain wall dynamics which
scales with $f^{3/2}H^{3/2}$) up to multidomain magnetichys-
teresis contribution. The later term can be explained us-
ing the Rayleigh correction to the magnetic permeability
and reveals a power loss scaling with $fH^3$. Curiously,
this type of behavior had been reported before in mag-
netic nanoparticle hyperthermia experiments \citep{HiergeistJMMMM99}. The
authors suggest that this can be explained by the exis-
tence of large particles in the sample \citep{HiergeistJMMMM99}. So, multi-
domain particles could be relevant to heat generation
through domain wall motion loss. However, for most
used magnetic fluid samples, multidomain particles are
not expected. For example, in magnetite nanoparticles
the single-domain limit is around 80 nm \citep{KrishnanIEEETM10}. Moreover,
on the theoretical point of view, Carrey et al. inves-
tigated the SLP field exponent using numerical simula-
tions of the SLLG equation (see Fig. 7 of Ref. \citep{CarreyJAP11}). The
authors found theoretically that this exponent is size de-
pendent and showed values below or higher than 2. This
type of behavior was found experimentally by Verde et
al. \citep{VerdeAIPAdv12}. However, in both works, no simple analytical
expression was used to explain this behavior.

Here we show that through a modification of Bloch's equation, which
is linear with respect to the magnetization, one is able to obtain
a heat loss expression valid beyond the LRT. Indeed,
different from other works from the literature, we demonstrate that
even the linear frequency term has higher order field contributions.
Also, our model introduces a nonlinear frequency term which adequately
describes the magnetic response within the hyperthermia therapeutic
window. The validity of the model is explicitly tested by comparing
it with numerical simulations of the SLLG approach. In addition, we
included experimental magnetic hyperthermia data that supports our
theoretical findings. Twelve powder samples were studied, including 
cobalt-ferrite, copper-ferrite, nickel-ferrite, maghemite and manganese-ferrite (doped
with Zn or Co and also undoped) based nanoparticles. 
The analytical nonlinear response model is believed
to be useful not only for improving our understanding of magnetic
losses, but also may impact other related areas, which could benefit
from analytical expressions, as for example, magnetic particle imaging\citep{GleichNature05,GoodwillAdvMat12}
and magnetic nanothermometry\citep{WeaverMedPhys09,ZhongSR14}.

The article is organized as follows: In section II we discuss several
models from the literature. In particular, we present the linear response
theory (LRT), the nonlinear Ferguson-Krishnan model (FK) (usually
applied in magnetic particle imaging), the perturbation method developed
by Raikher and Stepanov (RS model), and finally the stochastic Landau-Lifshitz-Gilbert
model, which is expected to be the exact solution of the magnetic
response of the nanoparticle at alternating field conditions. All
the models are critically compared showing the necessity of developing
a simple nonlinear analytical model. The SLLG model is than compared
with the proposed nonlinear response model (NLRT) developed in section
III. Section IV we present the experimental procedure, i.e. the synthesis
and characterization of magnetic fluids. In section V we discuss all
the theoretical and experimental results. Here we focus on magnetic
nanoparticle hyperthermia, but also compare our model with the third-harmonic
magnetic particle imaging data from the literature. Finally, in section
VI we summarize our findings.

\section{Models Review}

All the models discussed in this manuscript are valid
within the single-domain range. Also, they assume the giant-spin hypothesis
of Brown\citep{BrownPR63}, i.e. coherent spin rotation. Here we will
consider the case of uniaxial magnetic nanoparticle, where the energy
is given by 
\begin{equation}
E=KV\sin^{2}\theta-\mu_{0}M_{S}VH\cos(\theta-\varphi).
\end{equation}
The first term is the uniaxial anisotropy energy, while the other
is the Zeeman interaction. $\theta$ represents the angle between
the magnetic moment of the nanoparticle and the easy axis direction,
while $\theta-\varphi$ corresponds to the angle between the magnetic
dipole and the applied field. It is common to name the longitudinal
case as $\varphi=0$, which is the case where the field is applied
in the anisotropy axis direction.

The simplest quasi-static magnetization model in the
literature, named Langevin model, neglects the anisotropy term, which
can only be done if the ratio of this anisotropy contribution to the
thermal energy is very low. In this case the magnetization can be
calculated from 
\begin{equation}
\frac{M}{M_{S}}=\langle\cos\theta\rangle=\dfrac{\int\limits _{0}^{\pi}\cos\theta\mathrm{e^{\xi\cos\theta}}\sin\theta\mathrm{d}\theta}{\int\limits _{0}^{\pi}e^{\xi\cos\theta}\sin\theta\mathrm{d}\theta}=L(\xi).
\end{equation}
$L(\xi)=coth(\xi)-1/\xi$ is the Langevin
function, whose series expansion to fifth order gives 
\begin{align}
M_{\mathrm{}} & =M_{S}\left(\frac{\xi}{3}-\frac{\xi^{3}}{45}+\frac{2\xi^{5}}{945}-...\right)\\
 & =\chi_{LA,1}H+\chi_{LA,3}H^{3}+\chi_{LA,5}H^{5}+...\nonumber 
\end{align}
The first term is the initial (linear) susceptibility,
the second the cubic, and there on.

\subsection{Linear Response Theory}

The first linear response model to describe heat loss was probably
described by Debye in the context of rigid electric dipoles\citep{DebyePolMolec29}.
Here we focus in the magnetic case. Let us first start by assuming
that a magnetic particle is subjected to a harmonic field $H(t)=\Re\mathfrak{e}\left\lbrace H_{0}e^{i\omega t}\right\rbrace =H_{0}\cos{\omega t}$,
with the magnetic susceptibility $\chi=\chi'-i\chi''$, where $\chi'$
and $\chi''$ corresponds to the real and imaginary linear susceptibility
terms, respectively. So, the magnetization term can be written as
$M(t)=\Re\mathfrak{e}\left\lbrace \chi H(t)\right\rbrace =H_{0}\left(\chi'\cos{\omega t}+\chi''\sin{\omega t}\right)$,
where $\omega=2\pi f$ with $f$ the field frequency. Therefore, defining
the heating efficiency (SLP) as the frequency times the hysteresis
loss divided by the particle density ($\rho$) one finds 
\begin{align}
\label{SLPgeneric}
SLP & =\dfrac{f}{\rho V_{p}}\oint\limits _{\mathrm{cycle}}\delta Q=-\dfrac{f}{\rho}\mu_{0}\oint M\mathrm{d}H\\
 & =\pi\dfrac{f}{\rho}\mu_{0}H_{0}^{2}\chi''.\nonumber 
\end{align}
This equation represents the heat loss of the magnetic material. So,
one now needs an expression for the imaginary susceptibility term.
If the projection of the magnetization, $M(t)$,
in the field direction satisfies the Bloch equation, i.e. $\tau(\mathrm{d}M/\mathrm{d}t)+M=\chi H(t)$,
where $\tau$ is the magnetization relaxation time, one can show the
linear susceptibility term is $\chi={\chi_{0}}/{(1+i\omega\tau)}$,
revealing that 
\begin{equation}
\chi''=\chi_{0}\frac{\omega\tau}{1+(\omega\tau)^{2}}.
\label{ImchiLRT}
\end{equation}
$\chi_{0}$ is the equilibrium susceptibility, which
within the Langevin model is equal to $\chi_{LA,1}$. However, depending
on the model this term would be different. The relaxation of the
magnetization for an uniaxial nanoparticle is $\tau={\tau_{0}}e^{\sigma}/\sigma^{1/2}$
with $\sigma=KV/k_{B}T$, that is valid when $\sigma\geqslant2$
\citep{CoffeyJAP12}. Here $V$ is the particle volume, $T$ is the
temperature, $k_{B}$ in Boltzmann's constant and $K$ the magnetic
anisotropy. $\tau_{0}=\sqrt{\pi}M_{S}(1+\alpha^{2})/(\gamma_{0}2K\alpha)$
(about $10^{-10}-10{^{-8}}$s), with $\gamma_{0}$ the electron gyromagnetic
ratio and $\alpha$ the dimensionless damping factor. For the field
applied in the anisotropy direction one finds for the relaxation in
the limit of high anisotropy 
\begin{equation}
\tau_{h}=\frac{2\tau_{0}\left[\left(1-h\right)e^{-\sigma\left(1-h\right)^{2}}+\left(1+h\right)e^{-\sigma\left(1+h\right)^{2}}\right]^{-1}}{\sigma^{1/2}(1-h^{2})}.\label{taudeH}
\end{equation}
The field term $h$ is given in reduced units, i.e.
$h=H_{0}/H_{K}$. This expression returns to the later in the absence
of an applied field. The first one to describe this heat loss for
magnetic fluids was Rosensweig\citep{RosensweigJMMM02}. The model
above predicts a loss proportional to the square of the applied field.
However, this is only true experimentally at low field amplitudes
as found in several cases dealing with magnetic nanoparticles \citep{VerdeAIPAdv12,VerdeJAP12,HiergeistJMMMM99}.
Note that the same issue occurs in the electric case for dielectrics\citep{RichertPRL06}
or relaxor ferroelectrics\citep{GlazounovPRL00}. In addition, the
LRT model predicts elliptical magnetic hysteresis curves, which have
been observed at low field amplitudes (less than
$4~\mathrm{kA/m}$) by Eggeman et al. \citep{EggmanIEEETM07} and
Tomitaka et al. \citep{TomitakaJMMM12}. However, this is not consistent
with findings at higher field amplitudes, as for instance in magnetic
particle imaging where a nonlinear response plays a crucial role \citep{GleichNature05,GoodwillAdvMat12}.

\subsection{Ferguson-Krishnan Approach}

In an attempt to include nonlinear phenomena in the
description, Ferguson and Krishnan\citep{FergusonMedPhys11} proposed
a generalization of linear magnetization, using the
Langevin function: 
\begin{align}
M(t)= & M_{S}\left(\dfrac{1}{1+(\omega\tau)^{2}}L(\xi\cos(\omega t))\right.\\
 & +\left.\dfrac{\omega\tau}{1+(\omega\tau)^{2}}L(\xi\sin(\omega t))\right).\nonumber 
\end{align}
This approach assumes that the frequency response of higher field
order (quasi-static) terms are the same as the linear dynamic susceptibility
term and neglects the quasi-static contribution from the magnetic
anisotropy energy term. This expression is usually used to obtain
the nth-order harmonic magnetization, which represents an important
quantity in magnetic particle imaging \citep{GleichNature05,GoodwillAdvMat12}.
The harmonic calculation will be discussed in detail in section III.

\subsection{Raikher-Stepanov Perturbation Method}

Using perturbation theory, Raikher and Stepanov\citep{RaikherPRB97},
included the anisotropy term and showed that the magnetization could
be written as $M(t)=\Re(\chi_{1}H_{0}e^{i\omega t}+\chi_{3}H_{0}^{3}e^{3i\omega t}+...)$.
However, different from the FK model above, the frequency dependence
of the cubic term was found to be different from the linear term.
The authors found that the cubic susceptibility could be written as

\begin{equation}
\chi_{3}=-\dfrac{1}{4}\chi_{3}^{(0)}\dfrac{(1+S_{2}^{2})(1-i\omega\tau)}{45(1+i\omega\tau)(1+3i\omega\tau)},
\end{equation}
where $\chi_{3}^{(0)}=\phi\mu_{0}^{3}M_{S}^{4}V^{3}/(k_{B}T)^{3}$,
$\phi$ is the particle volume fraction of the assembly and $S_{2}=\dfrac{1}{2}\int\limits _{0}^{1}(3x^{2}-1)\mathrm{exp}(\sigma x^{2})\mathrm{d}x/\int\limits _{0}^{1}\mathrm{exp}(\sigma x^{2})\mathrm{d}x$.
So, the real and imaginary susceptibility terms are given by 
\begin{equation}
\chi_{3}'=\dfrac{1}{180}\chi_{3}^{(0)}\dfrac{(1+S_{2}^{2})(7\omega^{2}\tau^{2}-1)}{(1+\omega^{2}\tau^{2})(1+9\omega^{2}\tau^{2})},\label{eq:Rechi3RS}
\end{equation}

\begin{equation}
\chi_{3}''=-\dfrac{1}{180}\chi_{3}^{(0)}\dfrac{(1+S_{2}^{2})\omega\tau(3\omega^{2}\tau^{2}-5)}{(1+\omega^{2}\tau^{2})(1+9\omega^{2}\tau^{2})}.\label{eq:Imchi3RS}
\end{equation}

Using up to the cubic term the magnetization of the nanoparticle in
the RS model gives 
\begin{align}
M(t)= & \left(\chi_{1}'cos(\omega t)+\chi_{1}''sin(\omega t)\right)H_{0}\label{magRS}\\
 & +\left(\chi_{3}'cos(3\omega t)+\chi_{3}''sin(3\omega t)\right)H_{0}^{3},\nonumber 
\end{align}
where $\chi_{1}'=\chi_{1}^{(0)}(1+2S_{2})/(1+(\omega\tau)^{2})$,
$\chi_{1}''=\omega\tau\chi_{1}^{(0)}(1+2S_{2})/(1+(\omega\tau)^{2})$,
with $\chi_{1}^{(0)}=\phi\mu_{0}M_{S}^{2}V/(k_{B}T)$ and $\chi_{3}'$
and $\chi_{3}''$ are given by Eq. \ref{eq:Rechi3RS} and Eq. \ref{eq:Imchi3RS}.
Note that those expressions are valid for an ensemble and
low field amplitudes. In order to obtain the equivalent expressions
for the nanoparticle one only need to neglect the particle volume
fraction in the equilibrium susceptibilities. Because of the perturbation
approach this model is expected to be valid only at very low field
amplitudes.

\subsection{Stochastic Landau-Lifshitz-Gilbert Model}

The model that is expected to correctly describe the magnetization
response of a single-domain nanoparticle at any field
amplitude and frequency range is the SLLG model. In this case, the
magnetic moment of the nanoparticle is assumed to be described by
the Landau-Lifshitz-Gilbert equation 
\begin{equation}
\dfrac{\mathrm{d}\vec{M}}{\mathrm{d}t}=-\gamma\vec{M}\times\vec{H}_{\mathrm{eff}}-\dfrac{\alpha\gamma}{M_{S}}\vec{M}\times\left(\vec{M}\times\vec{H}_{\mathrm{eff}}\right),
\end{equation}
where 
\begin{equation}
\vec{H}_{\mathrm{eff}}(t)=\vec{H}_{\mathrm{}}(t)+\vec{H}_{\mathrm{ani}}+\vec{H}_{\mathrm{th}}(t).
\end{equation}
The effective field has three contributions: the applied external
field, the anisotropy field and the thermal fluctuation field. So
the Landau-Lifshitz-Gilbert equation for a magnetic dipole is augmented
with a Gaussian white noise thermal field $\vec{H}_{\mathrm{th}}$
whose Cartesian coordinates satisfy the statistical properties: $\langle\vec{H}_{\mathrm{th}}^{i}(t)\rangle=0$
and $\langle\vec{H}_{\mathrm{th}}^{i}(t)\vec{H}_{\mathrm{th}}^{i}(s)\rangle=2\left(k_{B}T\alpha/V\right)\delta_{ij}\delta(t-s)$.
The Kronecker and Dirac deltas indicates that the thermal field is
both spatial and temporally uncorrelated. In principle, one could
use the equation above and do numerical simulations. However, the
approach of Brown was to connect the SLLG equation to the Fokker-Planck
equation of the magnetic moment orientational distribution function
\citep{BrownPR63}, which can be used to obtain the nanoparticle magnetic
moment response.

In this work we focus in the longitudinal case. The
first authors to study in detail this problem analytically
was Dejardin and Kamilkov \citep{DejardinJAP09}. Later, others used
the same approach to describe dynamic magnetic hysteresis \citep{PoperechnyPRB10,LandiJAP12b,LandiPRB14}.
Here, the magnetic moment orientational distribution
function $f(z,t)$ can be shown to obey the Fokker-Planck equation
\begin{equation}
2\tau_{N}\dfrac{\partial f}{\partial t}=\dfrac{\partial}{\partial z}\left[\left(1-z^{2}\right)\left(\dfrac{\partial f}{\partial z}-f(z,t)h_{\mathrm{eff}}(z,t)\right)\right],
\end{equation}
with $\tau_{N}=\dfrac{\mu\left(1+\alpha^{2}\right)}{2\gamma_{0}\alpha k_{B}T}$
the free diffusion time, $z=\cos\theta$ and $\theta$
the angle between the magnetic dipole and the applied field. The magnetic
anisotropy is assumed uniaxial. So, the ratio of the particle energy
to thermal energy can be written as 
\begin{equation}
\dfrac{U_{\mathrm{eff}}}{k_{B}T}=-\sigma z^{2}-2h\sigma z,
\end{equation}
where the field term $h=H/H_{K}$. Therefore, the effective field
is 
\begin{equation}
h_{\mathrm{eff}}=-\dfrac{1}{k_{B}T}\dfrac{\partial U_{\mathrm{eff}}}{\partial z}=2\sigma(h+z).
\end{equation}

The Fokker-Planck equation is then used to obtain the time evolution
of the lth-order moment $p_{l}(t)=\langle P_{l}\rangle$, which is
shown to be described by 
\begin{equation}
2\tau_{N}\dfrac{\mathrm{d}p_{l}}{\mathrm{d}t}=\dfrac{l(l+1)}{2l+1}(A_{1}+A_{2})-l(l+1)p_{l},
\end{equation}
with 
\begin{equation}
A_{1}=2\sigma h(p_{l-1}+p_{l+1}),
\end{equation}
and 
\begin{align}
A_{2}= & 2\sigma\left[\dfrac{l-1}{2l-1}p_{l-2}+\dfrac{2l+1}{(2l-1)(2l+3)}p_{l}\right.\\
 & \left.-\dfrac{l+2}{2l+3}p_{l+2}\right].\nonumber 
\end{align}
This equation shows that each moment depends on others
in a nonlinear fashion. This infinite hierarchy may
be solved numerically using fast sparse solvers \citep{LandiJAP12b,LandiJAP12c,LandiPRB14,VerdeAIPAdv12}
and discarding several periods of the external field. Alternatively,
one could also expand the $p_{l}(t)$ in a Fourier
series as
\begin{equation}
p_{l}(t)=\sum_{k=-\infty}^{\infty}F_{k}^{l}(\omega)e^{ik\omega t},
\end{equation}
with all $p_{l}(t)$ real, which implies that $F_{-k}^{l}=(F_{k}^{l})^{*}$,
where the asterisks refer to the complex conjugate \citep{DejardinJAP09}.
This will then lead to a hierarchy of algebraic equations
for the Fourier amplitudes, which also need to be solved numerically\citep{DejardinJAP09}.

\subsection{Magnetization Loops}

\begin{figure}[htb]
\centering \includegraphics[width=1\columnwidth]{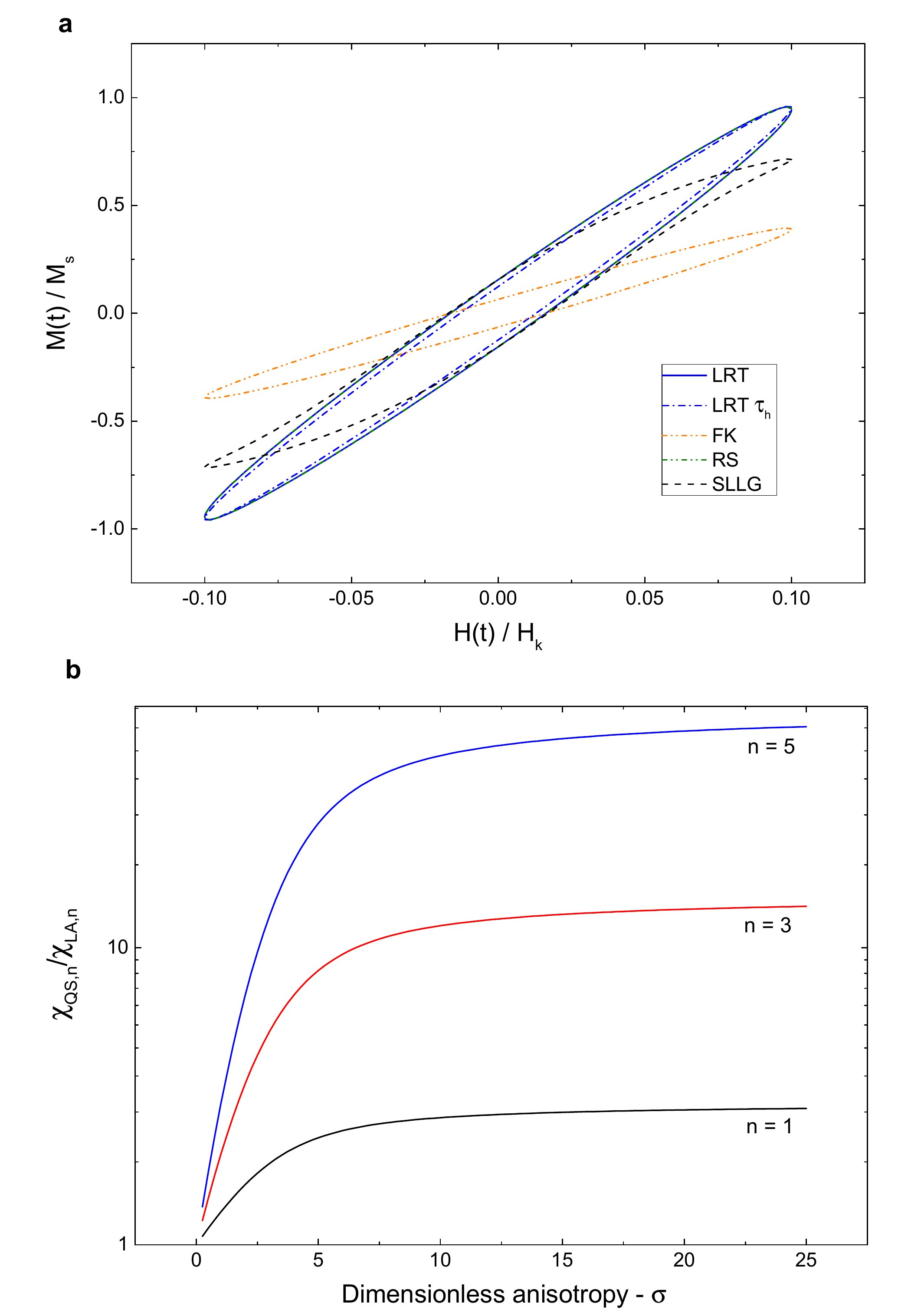} \caption{(Color online) (a) Dynamic hysteresis curves for the LRT, LRT considering
field dependence on relaxation time $\tau$, Ferguson-Krishnan approach,
Raikher Stepanov method and numerical solution of SLLG for $\sigma=6$
and $\omega\tau_{0}=10^{-3}$ . (b) Longitudinal to Langevin susceptibilities
ratio for $n=1$, $n=3$ and $n=5$.}
\label{figure2} 
\end{figure}

We are now in condition to compare the hysteresis loops of each model,
namely the linear response theory using the field-independent relaxation
time (LRT) and also the field-dependent relaxation time of Eq. \eqref{taudeH}
(LRT $\tau_{h}$), the FK model, the RS model and
the exact solution for the SLLG model. In Fig. \ref{figure2}(a) we
show the magnetization curves of all those models. It is clear
that the LRT model, independent of the relaxation time equation used,
shows an elliptical loop. The RS model showed a similar behavior.
The only model that shows a significant difference from LRT is the
FK model. However, it also shows an elliptical hysteresis, which is
distinct from the LRT model because of the Langevin equilibrium susceptibility.
So, different from the other models, it does not take into account
the anisotropy term. Nevertheless, for the parameters used in this
simulation, it is shown that none of the models above
represent well the exact solution given by the SLLG magnetization
hysteresis loop. Although improvements were obtained
in each model, in general they are not yet satisfactory.
Motivated by this fact, we decided to work out a nonlinear response
model, from now on named NLRT model. This model is able to better
represent the magnetization response at higher field conditions, not
only in comparison to the LRT, but also far better than the FK or
RS models.

\section{Theoretical Model}

In this section we present our nonlinear response model. Firstly,
we include the magnetic anisotropy energy term in the longitudinal
case, which allow us to obtain any quasi-static (equilibrium) susceptibility
terms. Those expressions will be named $\chi_{QS,n}$, i.e. the nth-order
quasi-static (QS) coefficient obtained in the low-frequency limit
($\omega\to0$). In the next subsection we introduce our dynamic model,
where a new expression for the heat loss and the particle magnetization
is obtained. The last subsection is related to the cubic harmonic
calculation, which is an important parameter for magnetic particle
imaging application.

\subsection{Quasi-static longitudinal case}

For an uniaxial magnetic nanoparticle in the longitudinal
case, the average magnetization is obtained from
\begin{equation}
\frac{M}{M_{S}}=\langle\cos\theta\rangle=\dfrac{\int\limits _{0}^{\pi}\cos\theta\mathrm{e^{\sigma\cos^{2}\theta+\xi\cos\theta}}\sin\theta\mathrm{d}\theta}{\int\limits _{0}^{\pi}e^{\sigma\cos^{2}\theta+\xi\cos\theta}\sin\theta\mathrm{d}\theta}.
\end{equation}
For $\sigma>0$, one can show that the longitudinal
magnetization is \citep{BakuzisJMMM01}

\begin{widetext}

\begin{equation}
M=M_{S}\left(\dfrac{2i\sinh{(\xi)}}{\sqrt{\sigma\pi}}\dfrac{e^{\sigma+\dfrac{\xi^{2}}{4\sigma}}}{\mathrm{erf}\left[i\left(\sqrt{\sigma}+\dfrac{\xi}{2\sqrt{\sigma}}\right)\right]+\mathrm{erf}\left[i\left(\sqrt{\sigma}-\dfrac{\xi}{2\sqrt{\sigma}}\right)\right]}-\dfrac{\xi}{2\sigma}\right).
\end{equation}

Expanding the longitudinal magnetization in a Taylor series:
\begin{align}
M_{\mathrm{}} & =M_{S}\left[\dfrac{ie^{\sigma}}{\sqrt{\sigma\pi}\mathrm{erf}(i\sqrt{\sigma})}-\dfrac{1}{2\sigma}\right]\xi+M_{S}\left[\dfrac{\mathrm{e}^{\sigma}\left(6\sigma\mathrm{e}^{\sigma}+i\sqrt{\sigma\pi}(2\sigma+3)\mathrm{erf}(i\sqrt{\sigma})\right)}{12\pi\left(\sigma\mathrm{erf}(i\sqrt{\pi})\right)^{2}}\right]\xi^{3}+...\\
 & =\chi_{QS,1}H+\chi_{QS,3}H^{3}+...,\nonumber 
\end{align}

\end{widetext}

where $\mathrm{erf}(iz)=(2i/\sqrt{\pi})\intop_{0}^{z}e^{u^{2}}du$
and $\chi_{QS,3}<0$. Note that all $\chi_{QS,n}$ are real. In
Fig. \ref{figure2}(b) we show the ratio of the $\chi_{QS,n}/\chi_{LA,n}$
up to the fifth-order (n=5). The longitudinal linear susceptibility
($\chi_{QS,1}$) calculation demonstrate
that in the absence of (or very low) magnetic anisotropy the susceptibility
approaches the expected Langevin result. On the other hand, in the
high anisotropy limit, the linear ratio approaches 3, which indicates
that the longitudinal result tends to the Ising result, as expected
in this case. Other ratios are also shown in the figure. Therefore,
we can conclude that in general it is of great importance to include
the anisotropy term when investigating the magnetic response of nanoparticles.

\subsection{Nonlinear Response Model}

As in LRT model, let us assume that a magnetic particle is subjected
to a harmonic field and that the projection of the
magnetization, $M(t)$, in the field direction satisfies the Bloch
equation, i.e. 
\begin{equation}
\tau\mathrm{\left(\frac{dM}{dt}\right)}+M=f(t),
\end{equation}
where $\tau$ is the relaxation time and $f(t)$ is a function of
the alternating field. Hence, it will be periodic, i.e. $f(t)=f(t+2\pi/\omega)$.
Also, in general one might represent $f(t)=\chi_{1}H(t)+\chi_{3}H(t)^{3}+...$,
where $\chi_{n}$ is the nth-order magnetic susceptibility. The LRT,
discussed before, corresponds to considering just the first term in
$f(t)$. The nonlinear response under Bloch's assumption may be computed
as follows. In general $f(t)$ is a function of $H(t)$ so it may
be expanded in a cosine series as $f(t)=\sum\limits _{n=1}^{\infty}c_{n}\cos(n\omega t)$
for certain coefficients $c_{n}$, that can be easily identified by
expanding $f(t)$ in terms of $cos(n\omega t)$ (another alternative
way to obtain those coefficients is using the integrating factor method
directly to Bloch's equation). The steady-state solution of the Bloch
equation is therefore 
\begin{equation}
M(t)=\sum\limits _{n=1}^{\infty}c_{n}\frac{\cos n\omega t+(n\omega\tau)\sin n\omega t}{1+(n\omega\tau)^{2}}.\label{magsumcn}
\end{equation}

In this approach, the corresponding SLP is 
\begin{align}
\text{SLP} & =\frac{\pi f}{\rho}H_{0}\frac{\omega\tau}{1+(\omega\tau)^{2}}c_{1}\label{eq4}\\
 & =\frac{\pi f}{\rho}\frac{\omega\tau}{1+(\omega\tau)^{2}}\left(\chi_{1}H_{0}^{2}+\frac{3}{4}\chi_{3}H_{0}^{4}+\frac{5}{8}\chi_{5}H_{0}^{6}+...\right).\nonumber 
\end{align}

This means that one only needs to worry with the coefficient $c_{1}(H_{0})$.
This comes from the fact that in the heat loss integral only the terms
obtained from $n=1$ is nonzero. Note that the first term corresponds
to the usual Debye model, if one assumes that $\chi_{1}=\chi_{QS,1}$,
i.e that $\chi_{1}$ is the quasi-static limit linear coefficient.
Also, it might be important to mention that the existence of the higher
order field dependent terms indicate a correction not reported before
in the literature. As for instance, if one uses the magnetization
equation of the RS model, only the the quadratic field term appears.
The same approach can also be used in the dielectric
loss case. For example, the electric field dependence dielectric loss
of glycerol (see inset of Fig. 3 of Ref \citep{RichertPRL06}).

According to equation \eqref{magsumcn}, the Bloch solution for the
magnetization $M(t)$ up to cubic terms in the field is 
\begin{align}
\label{magNLRT}
M(t)= & \left(\chi_{1}H_{0}+\dfrac{3}{4}\chi_{3}H_{0}^{3}\right)\dfrac{\cos(\omega t)+\omega\tau\sin(\omega t)}{1+(\omega\tau)^{2}}\\
 & +\dfrac{\chi_{3}H_{0}^{3}}{4}\dfrac{\cos(3\omega t)+3\omega\tau\sin(3\omega t)}{1+(3\omega\tau)^{2}},\nonumber
\end{align}
where $\chi_{n}$ are the nth-order magnetic susceptibility coefficients.
In the equation above is clear that higher-order
terms are also relevant to the magnetization dynamics. As for instance,
this nonlinearity effect can be identified even for the first harmonic
contribution, which shows higher field order terms.

In addition, if $\omega\tau\ll1$ one may write the magnetization
(considering higher-order terms in $f(t)$) as $M(t)=\chi_{1}H_{0}\cos(\omega t)+\chi_{3}H_{0}^{3}\cos(\omega t)^{3}+...$.
For the sake of argument, if one assumes that the nth-order susceptibility
terms are equal to the quasi-static terms ($\omega\to0$) and that
the nanoparticle is at the superparamagnetic regime, than $M(t)/M_{S}=L(\xi cos(\omega t))+\mathcal{O}(\omega\tau)$.
Note that the first term of this equation has been used systematically
in both, magnetic particle imaging (MPI)\citep{GleichNature05} and
magnetic nanothermometry (MNT)\citep{WeaverMedPhys09,ZhongSR14}.
In MNT the magnetization expression was shown to be useful only at
the low frequency range\citep{WeaverMedPhys09}, which is easily explained
by our model due to the range of validity of the later expression.
Moreover, in MPI the magnetization is
similar, but not identical to our model, and differs mainly due to
the term $n\omega\tau$ and that the later assumes quasi-static susceptibility
terms and superparamagnetic particle. As a consequence, our model
give different higher-order harmonic magnetization terms and might
represent better the experimental MPI data\citep{FergusonMedPhys11}.
Curiously, our model gives a similar expression as Ref. \citep{FergusonMedPhys11}
for the heat loss if we assume that $\chi_{n}=\chi_{QS,n}$. However,
this approximation does not represent correctly the magnetization
dynamics.

Further, Eq. \eqref{eq4} shows that the Bloch equation predicts the
same frequency dependence as the LRT, which will result
in elliptical-like hysteresis curves that are in disagreement with
experiment. The reason for this discrepancy is that Bloch's equation
is linear, whereas the underlying physical phenomena is not, as discussed
before in section II-D. One way to circumvent this is to assume that
the coefficients $\chi_{n}$ depend explicitly on $\omega$. The exact
form of this dependence is problem specific, but it must be such that
when $\omega\to0$, one recovers the equilibrium nonlinear susceptibilities.
The heuristic improvement approach, also used by others\citep{RaikherPRB97},
is able to better represent the magnetization dynamics.

So, to correct for the aforementioned deficiency of the Bloch approach,
we replace $\chi_{n}$ with a frequency dependent function and compare
the approximation with exact results, which are obtained for the longitudinal
case using the SLLG model\citep{VerdeAIPAdv12,VerdeJAP12,LandiJAP12c,CoffeyJAP12}.
In this strategy we wrote $\chi_{n}=\chi_{QS,n}g_{n}$, where $g_{n}$
is a function of the frequency. The quasi-static susceptibility coefficients
were obtained from the series expansion of the quasi-static longitudinal
solution\citep{BakuzisJMMM01}. Also, from our assumption is obvious
that one should have $g_{n}(\omega\tau\to0)=1$. Moreover, for the
first term we should have $g_{1}=1$, which corresponds to the LRT
result. For the cubic term we found that 
\begin{equation}
\chi_{3}=\chi_{QS,3}\dfrac{3-(\omega\tau)^{2}}{3\left(1+(\omega\tau)^{2}\right)}.
\label{chicomg}
\end{equation}

Similarly as the RS model, the magnetization can be
written in the same functional form as Eq. \eqref{magRS}. However, now the real
susceptibility terms are

\begin{equation}
\chi_{1}'=\dfrac{\chi_{QS,1}}{1+(\omega\tau)^{2}}+\dfrac{1}{4}H_{0}^{2}\chi_{QS,3}\dfrac{3-(\omega\tau)^{2}}{(1+(\omega\tau)^{2})^{2}},
\label{eq:Rechi3RS-1}
\end{equation}

\begin{equation}
\chi_{3}'=\dfrac{1}{12}\chi_{QS,3}\dfrac{3-(\omega\tau)^{2}}{(1+(\omega\tau)^{2})(1+(3\omega\tau)^{2})},
\label{eq:Rechi3RS-3}
\end{equation}
while the imaginary terms are $\chi_{1}''=\omega\tau\chi_{1}'$
and $\chi_{3}''=3\omega\tau\chi_{3}'$. Those results
indicate that the susceptibility terms are distinct from the RS model
(see Eqs. 10 and 11), even though the quasi-static susceptibility
coefficients give the same result. Also, the linear susceptibility
term shows a nonlinear field and frequency contribution, which was
absent in other models.

So, returning to the heat loss integral (Eq.
\eqref{SLPgeneric}) and using the cubic magnetization (Eq. \eqref{magNLRT}) with this correction (Eq. \eqref{chicomg}), the new expression for SLP is now given by 
\begin{align}
SLP= & \mu_{0}\pi\dfrac{f}{\rho}H_{0}^{2}\left[\dfrac{\chi_{QS,1}\omega\tau}{1+(\omega\tau)^{2}}\right.\label{SLPnlrt}\\
 & +\left.\dfrac{1}{4}H_{0}^{2}\dfrac{\chi_{QS,3}\omega\tau\left(3-(\omega\tau)^{2}\right)}{\left(1+(\omega\tau)^{2}\right)^{2}}\right].\nonumber 
\end{align}

Moreover, in section V, besides discussing the magnetic
nanoparticle hyperthermia, the cubic harmonic magnetic particle imaging
(MPI) experimental signal data obtained in Ref. \citep{FergusonMedPhys11}
will also be compared with the theoretical calculations using the
FK model and the NLRT model (see section V for details). We will
show a better agreement with experimental data using the nonlinear
response theoretical model developed in this work. In addition, because
we also investigate soft-magnetic nanomaterials (low $\sigma$), the
empirical uniaxial relaxation time expression, valid for any anisotropy
value, has been considered \citep{CoffeyJAP12}

\begin{equation}
\tau=\tau_{0}\left(e^{\sigma}-1\right)\left[2^{-\sigma}+\frac{2\sigma^{3/2}}{\sqrt{\pi}(1+\sigma)}\right]^{-1}.
\label{taudeH-1}
\end{equation}

\section{Experimental Procedure}

Manganese-ferrite samples were synthesized by hydrothermal route and
separated for the hyperthermia analysis after characterization by x-ray diffraction (XRD) and vibrating sample magnetometer (VSM).
All chemical reagents ($\mathrm{FeCl_{3}.6H_{2}O}$, $\mathrm{MnCl_{2}.4H_{2}O}$,
$\mathrm{ZnCl_{2}}$, $\mathrm{CoCl_{2}.6H_{2}O}$) citric acid trisodium
salt - $\mathrm{Na_{3}C_{6}H_{5}O_{7}}$, methylamine - $\mathrm{CH_{3}NH_{2}}$,
and acetone - $\mathrm{CH_{3}COCH_{3}}$) were purchased with analytical
quality and used without any further purification. In a typical approach,
$\mathrm{Mn_{0.75}[(Zn~or~Co)]_{0.25}Fe_{2}O_{4}}$ magnetic nanoparticles
were prepared as follows: adequate amounts of $1.0~\mathrm{mol/L}$
metal stock solutions were diluted with $40.0~\mathrm{mL}$ of distilled
water to form a precursor solution containing $10.0~\mathrm{mmol}$
of $\mathrm{Fe^{3+}}$, $3.75~\mathrm{mmol}$ of $\mathrm{Mn^{2+}}$,
and $1.25~\mathrm{mmol}$ of $\mathrm{Zn{}^{2+}}$or $\mathrm{Co^{2+}}$.
Thus, $120~\mathrm{mmol}$ of methylamine at 40\% (w/w) were quickly
poured into the stock solution under vigorous stirring for 10 min
and then transferred into a $120~\mathrm{mL}$ Teflon-sealed autoclave
and heated up to $160^{\circ}\mathrm{C}$ for 6 h. After cooling to
room temperature, the precipitate was separated by magnetic decantation,
washed with $\mathrm{H_{2}O}$ three times and re-dispersed in $50.0~\mathrm{mL}$
of water. Then, $4.0~\mathrm{mmol}$ of citric acid trisodium salt
was added into the solution which was heated up to $80^{\circ}\mathrm{C}$
for 60 min. After adjusting the pH of slurry to 7.0 and washing with
acetone three times, the precipitate was re-dispersed in $50.0~\mathrm{mL}$
of water to form a magnetic sol, after evaporating residual acetone.
Thus, a size-sorting process was done by adding $1~\mathrm{g}$ of
$\mathrm{NaCl}$ to the as-prepared magnetic sol\citep{SousaMicrochem11}.
5 min afterwards under a permanent magnet ($\mathrm{NdFeB}$), salt
adding induced a phase transition and formed an upper (botton) sol
phase with populations of smaller (larger) nanoparticles. Once separated,
precipitate of each phase was washed twice with a mixture water/acetone
1:10 (volume/volume) and, after evaporating residual acetone, nanoparticles
were re-dispersed in water. This procedure was repeated several times.
Powders were obtained from evaporation of sols at $55^{\circ}\mathrm{C}$
for 8 h. Details about cobalt-ferrite samples can be found in Ref. \citep{VerdeJAP12} and copper-ferrite and nickel-ferrite samples can be found in Ref. \citep{VerdeAIPAdv12}.

After the size-sorting process powder samples were analyzed by XRD
(Shimadzu 6000) to separate samples with similar sizes. The previous
analysis was performed using the well-known Scherrer equation,
which is given by $D_{XRD}=\kappa\lambda/\beta \cos\psi$, where $\kappa=0.89$
is the Scherrer constant, $\lambda=0.15406~\rm{nm}$ is the X-ray wavelength,
$\beta$ is the line broadening in radians obtained from the square
root of the difference between the square of the experimental width
of the most intense peak to the square of silicon width (calibration
material), and $\psi$ is the Bragg angle of the most intense peak
(311). This procedure allowed us to select three distinct samples
of similar sizes containing $\mathrm{MnFe_{2}O_{4}}$,
$\mathrm{Mn_{0.75}Zn_{0.25}Fe_{2}O_{4}}$, or $\mathrm{Mn_{0.75}Co_{0.25}Fe_{2}O_{4}}$
nanoparticles. All the nanoparticles were surface-coated with citric
acid, which guarantee stability at phisiological conditions. The
samples were also characterized by VSM (ADE Magnetics, model EV9,
room temperature measurements, field up to 2T). Table \ref{table1}
summarizes the relevant characterization properties of the nanoparticles.

Finally, magnetic hyperthermia data was obtained in two systems, 
one home-made which operates at 500kHz, and another one from nanoTherics. 
In particular, the later system operates in a
broad frequency range, spanning from 110 up to 980kHz. While details 
about the home-made hyperthermia system has been described elsewhere
\citep{VerdeJAP12,VerdeAIPAdv12}. The calorimetric
method used to obtain the experimental SLP of the sample used the
equation
\begin{equation}
SLP=\frac{C}{m_{NP}}\left[\frac{dT}{dt}\right]_{max},\label{chicomg-1-1-1}
\end{equation}
where $C$ is the heat capacity of the sample (here
assumed as the heat capacity of the liquid carrier due to the low
concentration of particles), $m_{NP}$ is the mass of magnetic nanoparticles
in unit of grams (obtained from the analysis of the magnetisation
curves of the colloid samples), $T$ is the temperature of the sample
measured with a fibre optic thermometer. Note that in the SLP calculation
we use the value of the maximum rate of temperature increase ($[dT/dt]_{max}$),
as discussed previously by others \citep{VerdeAIPAdv12,BordelonJAP11}. This method is believed to better estimate SLP than
the most common initial-slope procedure that can underestimate this
value \citep{AndreuIJH13}.

\section{Results and Discussion}

\subsection{Theoretical results}

\begin{figure}
\centering \includegraphics[width=\columnwidth]{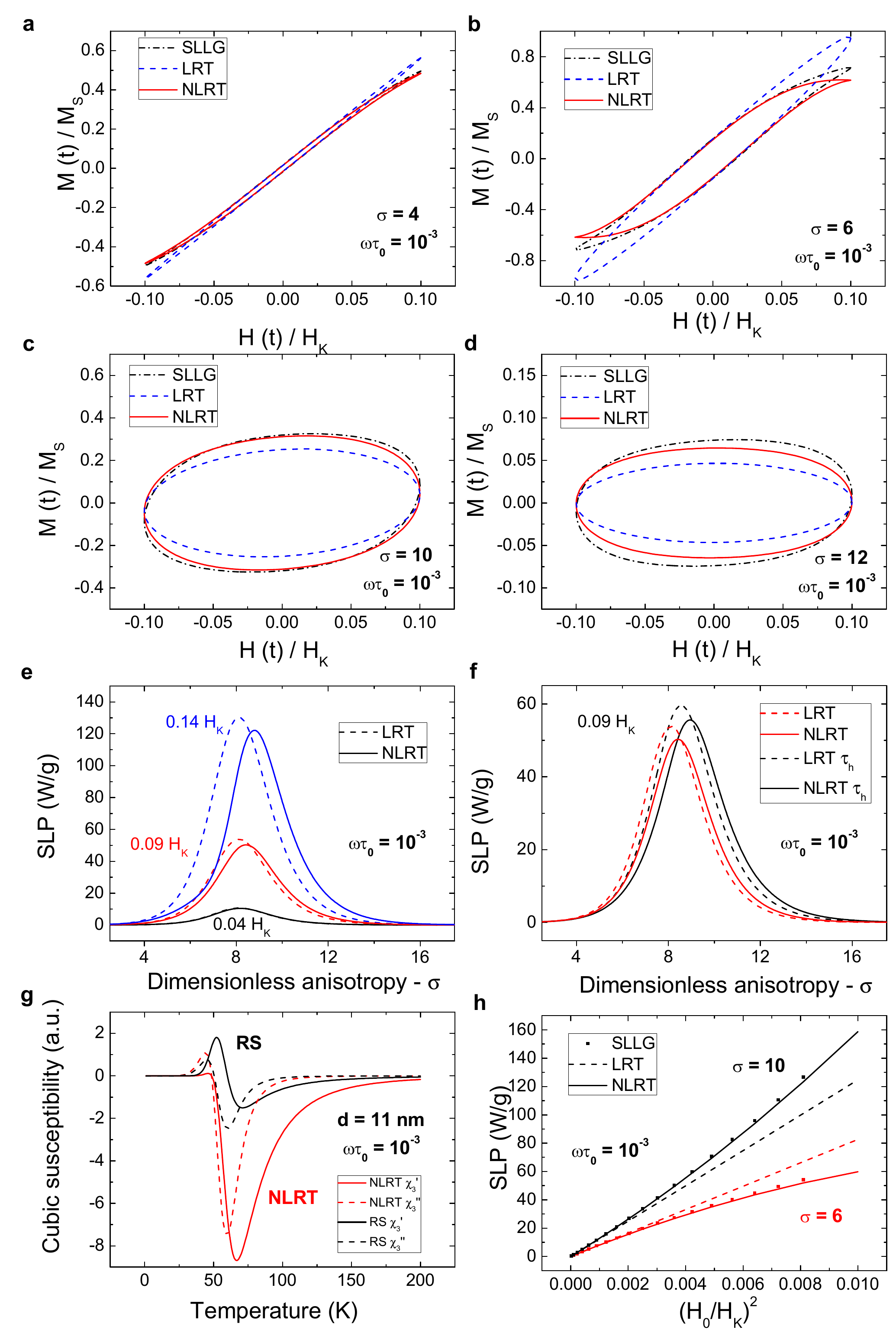} \caption{(Color online) Dynamic hysteresis curves for the LRT (dashed line),
NLRT (solid line) and the exact solution (dash-dot line) using the
SLLG equation for field $H_{0}=0.1H_{K}$ and $\omega\tau_{0}=10^{-3}$
. In (a) $\sigma=4$, (b) $\sigma=6$, (c) $\sigma=10$ and (d) $\sigma=12$.
(e) SLP as function of $\sigma$ for the LRT and NLRT with distinct
field amplitudes. (f) SLP as function of $\sigma$ for the LRT and
NLRT with and without the field dependence on the relaxation time.
(g) Real and imaginary susceptibilities as function of temperature
for the RS and the NLRT models. (h) SLP as a function the square of
the field for the LRT (dash), NLRT (solid) and exact solution using
the SLLG (points) for $\sigma$ values of 6 and 10.}
\label{figure3} 
\end{figure}

Several experimental results show the existence of an optimal particle
size for hyperthermia \citep{CarreyJAP11,KrishnanIEEETM10,VerdeJAP12}.
This is also contemplated in Eq. \eqref{ImchiLRT}, which predicts that this optimal
size should occur when $\omega\tau=1$. This, however, is only true
at low field amplitudes. Increasing field amplitude one notice a shift
of maximum size towards larger particles in a noninteracting system.
This can be easily modelled within LRT using the field dependent magnetization
relaxation time \citep{DejardinJAP09}. Indeed, such drift becomes
clear when $h>0.04$ (see discussion
of Fig. \ref{figure3}(f) below). Further, numerical dynamic hysteresis
simulations using the SLLG model or Kinetic Monte Carlo method \citep{VerdeAIPAdv12,VerdeJAP12,RutaSR15,CarreyJAP11}
show that, as the field amplitude increases, the optimal size shifts
towards larger particles. It may even disappear, depending (also)
on the magnetic anisotropy of the nanoparticle \citep{CarreyJAP11,VerdeAIPAdv12,VerdeJAP12,RutaSR15}.
Most of the results above consider a noninteracting system. However,
in colloids, or real in vivo situation, agglomerate formation plays
a key role. In this case, it has been shown within LRT, that the opposite
effect occurs, i.e. increasing the strength of particle interaction
shift the optimal diameter to lower sizes \citep{BranquinhoSR13}.
The same was found including particle-particle interaction using a
mean field approach to the SLLG model at the low field regime \citep{LandiPRB14}. Anyway,
a valuable analytical nonlinear response theoretical model (NLRT)
should be able to explain at least some of the features discussed
above. 

A comparison between the hysteresis curves obtained from the LRT,
our NLRT model and the numerical solution of the SLLG model is shown
in Fig.\ref{figure3}(a)-(d), for distinct $\sigma$ values considering
$\omega\tau_{0}=10^{-3}$ and $H_{0}/H_{K}=0.1$. It is found that
the inclusion of the corrected cubic term leads to a good agreement
with the numerical simulations, adequately describing the deviations
from the linear response. Note that the agreement
is far better than any other model discussed previously (see Fig.
\ref{figure2}(a)). The LRT model is shown as a dash line, the exact
result using the SLLG equation is shown as dash-dot line, while NLRT
(considering Eq. \eqref{chicomg}) is shown as solid line. It is very
surprising that, with such a simple assumption, an interesting nonlinear
effect is obtained able to represent far better the magnetization
dynamics. Indeed, we found that the present model works very well
close to this limit of anisotropy value ($H_{0}\cong0.14H_{K}$).
It also has a slight frequency dependence which can be monitored by
nonphysical results in the magnetization curve or kinks in the SLP versus
$\sigma$ curves increasing the field. At higher fields
we observe deviations from the exact solution that might be only addressed
if higher-order terms are determined. Nevertheless, as shown in Fig.
\ref{figure1} (see the NLRT line), the range of validity of the model
is almost completely within the hyperthermia therapeutic window. This
suggests that this model might be applicable for real clinical situations.

Figure \ref{figure3}(e) shows the SLP as function of $\sigma$
for the LRT (dashed line) and the NLRT (solid line) for distinct field
amplitudes. For simplicity, we are not considering the field dependence
on the relaxation time. One can clearly observe a
shift of the maximum of SLP towards higher particle sizes in the nonlinear
case. Also a decrease of the maximum SLP value for the NLRT case.
The phenomenon is strictly related to the nonlinear effect introduced
in the model and not due to the field effect from the relaxation.
This result is in accordance with numerical simulations from the literature \citep{VerdeJAP12,RutaSR15}.
On the other hand, Fig. \ref{figure3}(f) also shows SLP as
function of $\sigma$ in both cases, but now investigating the field
effect on the relaxation time for $H=0.09H_{K}$. Similar behavior
as before is observed. Nevertheless, in comparison with the LRT, the
NLRT-$\tau(H)$ shows a larger size shift. As for instance, the optimum
anisotropy term change from $\sigma_{opt}=8.1$ for LRT to $\sigma_{opt}=9.0$
for NLRT-$\tau(H)$, which corresponds to a shift in optimal diameter
of the order of 4\%.

As discussed in section II, there are other nonlinear models. In particular,
cubic susceptibility expressions using the RS model had been suggested
to represent experimental data of noninteracting magnetic nanoparticles \citep{JonssonJMMM00}.
Figure \ref{figure3}(g) shows the cubic susceptibility terms,
imaginary and real, for the RS model and the NLRT model as a function
of temperature. Here the parameters used were $d=11~\mathrm{nm}$,
$M_{S}=270~\mathrm{emu/cm^{3}}$, $K_{ef}=8\times10^{4}~\mathrm{erg/cm^{3}}$,
$\alpha=0.05$ and $\rho=5~\mathrm{g/cm^{3}}$. As found in Ref. \citep{RaikherPRB97}
the real cubic term in the RS model shows a significant variation
as a function of temperature, in particular in the range below $60~\mathrm{K}$,
where a quite high positive cubic value is found theoretically. It
is curious to notice that experimentally such effect has not been
observed in Ref. \citep{JonssonJMMM00} for noninteracting nanoparticle
samples. In fact, discrepancies between the RS model and data of Ref.
\citep{JonssonJMMM00} had been attributed to polydispersity and particle-particle
interaction effects. Note that the inclusion of such
effects could be responsible for some of those differences between
theory and data. However, there might be another explanation. As we
have just shown, the NLRT model represents far better the magnetization
response. Differently from the RS model, the real cubic susceptibility
from NLRT does not show such strong positive contribution at low temperatures.
As a consequence it might represent better experimental data. Another
point that could be commented about the improvement in the NLRT model
in comparison to others is the SLP calculation. Note that in the RS
model the SLP calculation, using the magnetization expression of Eq. \eqref{magRS},
provides the same result as the LRT. So, although the magnetization
equations are not the same, the hysteresis area is the same as the
LRT case. Again, this is in contradiction with several experimental
results. From the experimental point of view, after obtaining
the SLP data of the samples as a function of the applied alternating
field, it is common to try to describe the heating efficiency in terms
of a field exponent, i.e. one might try to fit the data with an allometric
expression as $SLP=aH^{\nu},$ where $a$ is a constant and $\nu$
the field exponent. If this exponent is equal to 2 one might argue
that the sample is within the linear response regime. 

Figure 3(h) shows the SLP as a function of the quadratic
field for distinct $\sigma$ values considering the LRT (dash), NLRT
(solid) and SLLG (points). Both situations shows that depending on
the particle size or anisotropy, deviations from the expected quadratic
field dependence of the LRT are found. At the low barrier regime ($\sigma<\sigma_{opt}$),
i.e. for particle sizes lower than the optimum value, the field dependence
exponent is lower than 2. While at the high barrier regime, an exponent
higher than 2 is observed. The same behavior is found
from SLLG, as expected since NLRT model is based on
the assumption that SLLG is the exact result. However, because in
the NLRT only the cubic term was introduced, deviations between both
models are expected for higher fields. The nonlinear regime has been
studied experimentally before on Ref. \citep{VerdeAIPAdv12}, where
the transition to the nonlinear regime was explained using the SLLG
model, though without any analytical expression. The explanation for
such behavior may be understood using Eq. \eqref{magNLRT}. Note that $\chi_{QS,3}<0$,
so when $\omega\tau<\sqrt{3}\approx1.7$ the high-order contribution
term lowers the linear SLP field dependence term. The consequence
of this is an apparent field exponent lower than 2. On the other hand,
when $\omega\tau>\sqrt{3}$ the higher order SLP term changes sign,
which now adds a value to the first order term. In this case exponents
larger than 2 might appear if the field is high enough.

\subsection{Magnetic hyperthermia evidence}

Evidence of nonlinear behavior the SLP field dependence can be found in
distinct ferrite-based powder samples. Table \ref{table1} summarizes the parameters
obtained from sample analysis. Four sets of samples were
studied. The first set is composed of three samples: manganese-ferrite based
nanoparticles undoped, doped with zinc and doped with cobalt. Since samples 
were produced using the same method, have (approximately) the same magnetization,
and the same diameter, this set allow the study of anisotropy influence over SLP 
versus $H$ behavior. The second set is composed by other three samples:
cobalt-ferrite, maghemite and copper-ferrite. These samples have very different
magnetization and anisotropy, but the same diameter (some results published in Ref. \citep{VerdeJAP12}). The third set is composed by
other three samples of cobalt-ferrite, which have a high anisotropy,
with different diameters. And, the last set is composed by three samples of 
nickel-ferrite, which have a lower anisotropy than cobalt-ferrite, with different diameters.

\begin{table} \centering \begin{tabular}{c c c c c c c c c c c c}
\hline \hline
Sample & $D_{\mathrm{XRD}}$ & $M_{S}$ & $H_{\mathrm{coer}}$ & {$\nu$} \\
 & (nm) & (emu/cm$^3$) & (Oe) & \\ \hline
$\mathrm{MnFe_{2}O_{4}}$ & 11.3 & 293 & 21 & 2.2 \\
$\mathrm{Mn_{0.75}Zn_{0.25}Fe_{2}O_{4}}$ & 11.1 & 302 & 0.4 & 1.6 \\
$\mathrm{Mn_{0.75}Co{}_{0.25}Fe_{2}O_{4}}$ & 11.4 & 309 & 77 & 2.6 \\ 
\hline
$\mathrm{CoFe_{2}O_{4}}$ & 9.1 & 272 & 152 & 3.9 \\
$\mathrm{\gamma-Fe_{2}O_{3}}$ & 9.3 & 209 & 2.7 & 2.0 \\
$\mathrm{CuFe_{2}O_{4}}$ & 9.4 & 124 & 0.5 & 1.2 \\ 
\hline
$\mathrm{CoFe_{2}O_{4}}$ & 3.4 & 103 & 1.4 & 1.9 \\
$\mathrm{CoFe_{2}O_{4}}$ & 12.9 & 253 & 261 & 2.5 \\
$\mathrm{CoFe_{2}O_{4}}$ & 13.6 & 281 & 299 & 5.5 \\ 
\hline
$\mathrm{NiFe_{2}O_{4}}$ & 5.3 & 153 & 0.3 & 1.5 \\
$\mathrm{NiFe_{2}O_{4}}$ & 7.9 & 151 & 0.4 & 2.1 \\
$\mathrm{NiFe_{2}O_{4}}$ & 12.8 & 185 & 4.4 & 2.3 \\ 
\hline \hline
\end{tabular}
\caption{Characterization parameters of the samples. $D_{\mathrm{XRD}}$ crystalline size, $M_{S}$ saturation magnetization and $H_{\mathrm{coer}}$ coercive field. {$\nu$} is the apparent SLP field exponent from allometric fit.}
\label{table1}
\end{table}

\begin{figure}
\centering \includegraphics[width=0.95\columnwidth]{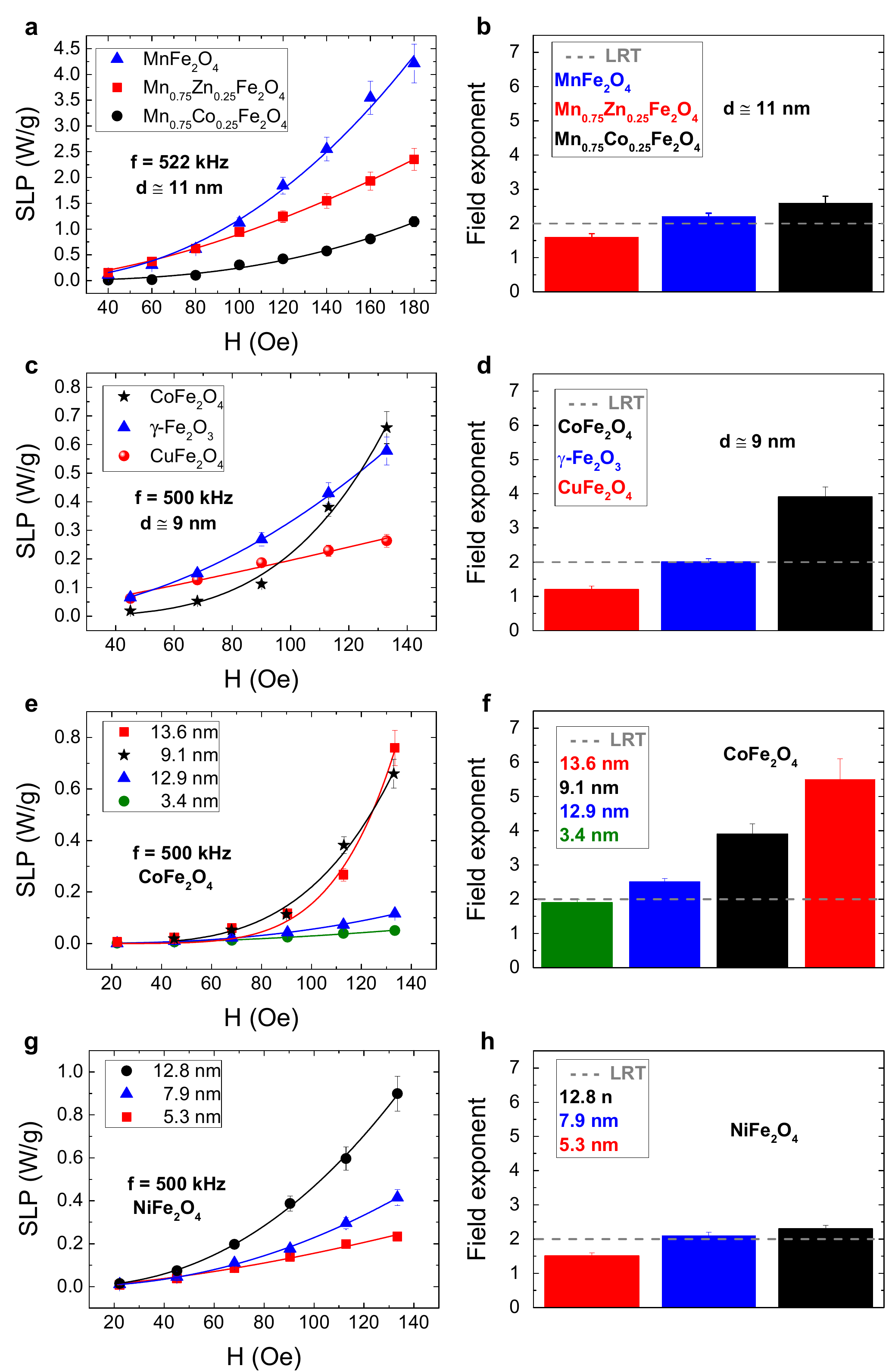} \caption{(Color online) (a) SLP as function of the magnetic
field for distinct manganese-ferrite nanoparticles around $11~\rm{nm}$ in powder configuration
at $f=522$ kHz. (b) Apparent SLP field exponent $\nu$
obtained for manganese-ferrite in powder configuration.
(c) SLP as function of the magnetic
field for distinct ferrite nanoparticles around $9~\rm{nm}$ in powder configuration
at $f=500$ kHz. (d) Apparent SLP field exponent $\nu$
obtained for distinct ferrite nanoparticles in powder configuration.
(e) SLP as function of the magnetic
field for cobalt-ferrite nanoparticles with distinct sizes in powder configuration
at $f=500$ kHz. (f) Apparent SLP field exponent $\nu$
obtained for cobalt-ferrite in powder configuration.
(g) SLP as function of the magnetic
field for nickel-ferrite nanoparticles with distinct sizes in powder configuration
at $f=500$ kHz. (h) Apparent SLP field exponent $\nu$
obtained for nickel-ferrite in powder configuration.
Symbols are data and lines represent the best fit
using the allometric function.
}
\label{figure4} 
\end{figure}

Magnetic hyperthermia experimental data around $500~\mathrm{kHz}$ is
shown in Figs. \ref{figure4}(a), \ref{figure4}(c), \ref{figure4}(e) and
\ref{figure4}(g) for powder samples, where we present
the SLP as a function of the applied field for distinct ferrite-based
samples. Most of the applied fields are above the therapeutical values
(see Fig.\ref{figure1}), but are necessary to experimentally
observe deviations from LRT. Symbols represent experimental data,
while the lines are the fit of the data using the allometric function.
Firstly, notice that soft-like materials heat more
efficiently at low field amplitudes, in aggreement with what was found
before experimentally and theoretically \citep{BranquinhoSR13,VerdeJAP12,VerdeAIPAdv12}.
This property, although not discussed in this work, is relevant for
in vivo applications \citep{RodriguesIJH13}. Table
\ref{table1} shows the apparent field exponents
obtained from this type of phenomenological approach for all the samples,
as well Figs. \ref{figure4}(b), \ref{figure4}(d), \ref{figure4}(f) and
\ref{figure4}(h), compared with 2 (gray dashed line which represents LRT).
The result indicates deviation from linear behavior and the samples shows
distinct exponents values, depending (probably) on sample anisotropy. The same behavior has
been observed with other ferrites \citep{VerdeAIPAdv12}. This behavior
is in accordance with our previous theoretical analysis. However, a direct comparison
between experimental data and theoretical analysis is compromised by the fact that
sample are solid, allows a random anisotropy axis nanoparticle
configuration that decreases the equilibrium susceptibility values
lowering the SLP \citep{VerdeAIPAdv12}. So, the
nanoparticles at this highly packed configuration are at strong interacting
conditions, which may affect the magnetic anisotropy \citep{VerdeAIPAdv12,BranquinhoSR13}.
In this case, one can not use the longitudinal calculation
developed in this work for the powder samples, since the quasi-static
susceptibility values are now different. Nevertheless, powder configuration inhibit frictional loss
contributions due to the Brownian relaxation mechanism \citep{RaikherJMMM08,RaikherPRE11,ShubitidzeJAP15} and a similar behavior
for SLP (with distinct absolute values) is also expected.

The NLRT model developed here is valid
for $H\leq0.14H_{K}$, where magnetization relaxation mechanisms plays
a role in the spin reorientation by overcoming the barrier energy.
Increasing the field value one need to use directly the SLLG model,
which due to the complexity of the problem does not reveal any simple
analytical equation. Nevertheless, a simple approach for qualitative
analysis under high field conditions ($H>H_{K})$ might be achieved
using the Stoner-Wohlfarth (SW) model \citep{HergtIEEETM98}.

\begin{figure}[htb]
\centering \includegraphics[width=1\columnwidth]{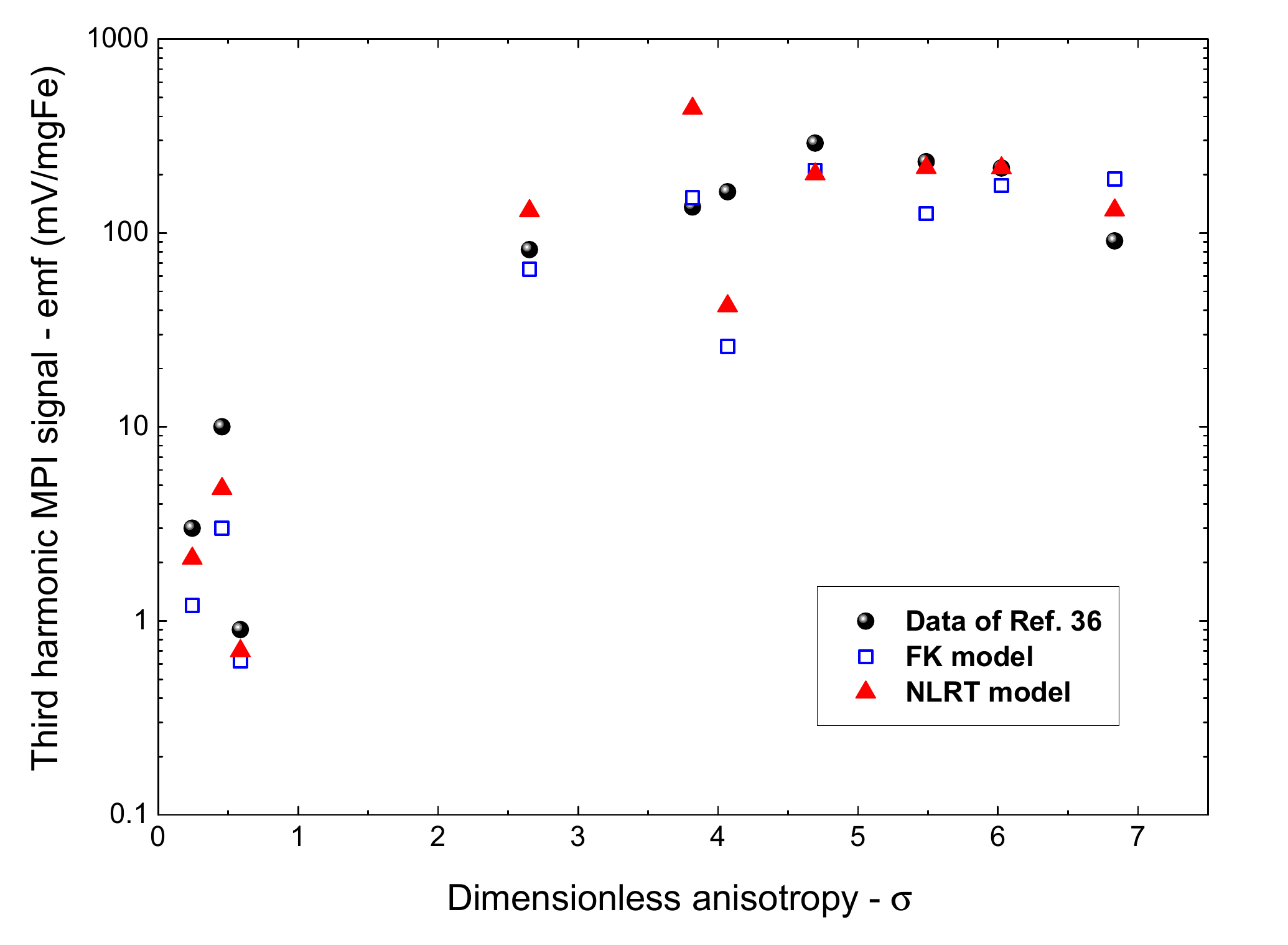} \caption{(Color online) MPI third harmonic signal of magnetite-based
magnetic fluids containing nanoparticles of different sizes as a function
of $\sigma$. The figure shows the experimental data (circles) from
Ref. \citep{FergusonMedPhys11}, calculations using the FK model of
Ref. \citep{FergusonMedPhys11} (squares) and the NLRT (triangles)
calculation.}
\label{figure5} 
\end{figure}

\subsection{Magnetic particle imaging evidence}

Besides magnetic hyperthermia, the present model might be useful for 
magnetic particle imaging (MPI) too. MPI is a nonionizing imaging
technique, introduced in 2005 by Gleich and Weizenecker \citep{GleichNature05},
which is capable of imaging magnetic tracers through the nonlinear
magnetic response of magnetic nanoparticles. In MPI a DC plus an AC
field are applied to the magnetic material in such a way to create
a free field point volume where the nanoparticles can respond to the
ac field excitation. The magnetic response signal can then be measured
using detector coils. The received voltage by the detector coil is
\begin{equation}
u=-\mu_{0}\int\limits _{V}S_{0}(x)\dfrac{\partial M(x,t)}{\partial t}\mathrm{d}V,
\label{MPIsignal}
\end{equation}
$S_{0}$ is the coil sensitivity (assumed to be $\mu_{0}S_{0}=2.25~\mathrm{mT/A}$)
and the integration is over the magnetic material. The MPI third harmonic
magnetization signal per unit volume $\mathrm{\mathbf{emf}}_{3\omega_{0}}$
is defined as the module of the Discrete Fourier Transform given by
\begin{equation}
\mathrm{\mathbf{emf}}_{3\omega_{0}}=\mu_{0}S_{0}\left|DFT[u_{3}]\right|,\label{MPIemf-1-1}
\end{equation}
where 
\begin{equation}
DFT[u_{3}]=\sum_{k=0}^{N-1}f\left[k\right]e^{-i\frac{6\pi}{N}k}.
\end{equation}
The function $f\left[k\right]$ is obtained using $f\left[t\right]=\frac{\partial M(t)}{\partial t}$
and the time discretization as $t=\frac{k}{Nf_{0}}$, where $f_{0}$
is the excitation field frequency and $N$ corresponds to the number
of intervals discretized within one period. In this work $N=40$.
In NLRT model the complete magnetization expression is unknown, so
we only use the terms up to the third harmonic. On the other hand,
for the FK model, one can expand the Langevin expression up to any
order.

Figure \ref{figure5} we shows the experimental
MPI data of the third harmonic magnetization signal of magnetite nanoparticles
of distinct sizes performed at $250~\mathrm{kHz}$ (see Ref. \citep{FergusonMedPhys11}
for details). Spheres correspond to experimental data, while squares
are related to the FK model of Ref. \citep{FergusonMedPhys11}. Note
the logarithmic scale and that we are presenting the data in terms
of $\sigma$. Here we assumed the bulk magnetic anisotropy value,
although is well known that the anisotropy is size dependent \citep{BakuzisJMMM01,BodkerPRL94,BakuzisJMR96,BakuzisJAP99}.
Nevertheless, size dispersity was taken into account. The calculations
used a relaxation time valid for
any $\sigma$ \citep{CoffeyJAP12,BranquinhoSR13} and parameters from
Table 1 of Ref. \citep{FergusonMedPhys11}. Triangles
correspond to our polydisperse calculation taking the Discrete Fourier
Transform and using Eq. \eqref{MPIsignal} in
units of V/g, i.e. taking into account in the calculation of $\mathrm{\mathbf{emf}}_{3\omega_{0}}$
the amount of magnetic material in mass per unit volume. Note that our model represents
better the MPI experimental data. Indeed from 10 data points NLRT
is in better agreement with 80\% of the data. Better theoretical results
might be achievable if the anisotropy of each sample is known, or
even more if one is able to take into account possible particle-particle
interaction effects due to agglomerate formation \citep{BakuzisACIS13}.
So, it might be fair to say that, both hyperthermia and MPI experiments
seem to be more adequately described by the NLRT model.

Finally, it might be relevant to comment that there is a huge interest
of not only deliver heat using magnetic nanoparticle hyperthermia, but also,
monitor non-invasively heat delivery using magnetic nanoparticles. In order
to be successful in such area, analytical expressions, as the ones derived
in this work, that better represent the non-linear response of magnetic
nanoparticles, are highly needed. The authors believe that the model
developed here might indicate a useful approach towards this important
clinical goal. 

\section{Conclusion}

In conclusion, a nonlinear response model of magnetic nanoparticles
valid for single-domain nanoparticles was developed.
The model is valid beyond the linear response theory, and showed good
agreement with dynamic hysteresis simulations using the stochastic
Landau-Lifshitz-Gilbert approach and experimental hyperthermia data
for field amplitudes as high as 10\% of the magnetic anisotropy field.
In particular, a generalized expression for the magnetization and
the heat loss efficiency (SLP) were obtained. The model showed many 
features found experimentally in magnetic
hyperthermia and MPI studies. As for example, Stoner-Wohlfarth-like
dynamic hysteresis curves, distinct SLP field exponents, optimum hyperthermia
nanoparticle size shift, among others. The magnetization expression 
was critically compared with the ones used in MPI and MNT, from which
we were able to identify when some approximations can be used.
Moreover, the NLRT was found to be valid mostly within
the hyperthermia therapeutic window, which suggests strong applicability
in the biomedical field.

\acknowledgments{The authors would like to thank financial support
from the Brazilian agencies CNPq, CAPES, FAPEG, FAPESP (2014/01218-2),
FAPDF and FUNAPE.}

%
\end{document}